\documentclass[preprint,showpacs,amsmath,amssymb,superscriptaddress]{revtex4}
\usepackage{amsmath,amssymb,color,latexsym,natbib,graphicx}

\usepackage{dcolumn}
\usepackage{bm}
\newcommand{\p} {\partial}
\def\eg{{\it e.g.}\ } 
\def\ie{{\it i.e.}\ }
\def\uu{{\bf u}}
\def\rr{{\bf r}}
\def\be{\begin{equation}}
\def\ee{\end{equation}}
\def \pmbmath{\mathpalette\pmbmathaux}
\def \pmbmathaux#1#2{
         \pmbtext{$#1#2$}}
\def \pmbtext#1{\leavevmode
     \setbox0\hbox{#1}
     \kern0,4pt \copy0 \kern-\wd0
     \kern-0,2pt \raise0,3pt \box0 }
     
\begin{document}

\preprint{1}

\title{Exact relation for correlation functions in compressible isothermal turbulence}
\author{S\'ebastien Galtier}
\affiliation{Univ Paris-Sud, Institut d'Astrophysique Spatiale, UMR 8617, b\^at. 121, F-91405 Orsay, France}
\affiliation{Institut universitaire de France}
\author{Supratik Banerjee}
\affiliation{Univ Paris-Sud, Institut d'Astrophysique Spatiale, UMR 8617, b\^at. 121, F-91405 Orsay, France}

\date{\today}
\begin{abstract}
Compressible isothermal turbulence is analyzed under the assumption of homogeneity and in the asymptotic 
limit of a high Reynolds number. An exact relation is 
derived for some two-point correlation functions which reveals a fundamental difference with the incompressible 
case. The main difference resides in the presence of a new type of term which acts on the inertial range similarly 
as a source or a sink for the mean energy transfer rate. When isotropy is assumed, compressible turbulence may 
be described by the relation, $- {2 \over 3} \varepsilon_{\rm{eff}} r = {\cal F}_r(r)$, where ${\cal F}_r$ is the radial 
component of the two-point correlation functions and $\varepsilon_{\rm{eff}}$ is an effective mean total energy 
injection rate. By dimensional arguments we predict that a spectrum in $k^{-5/3}$ may still be preserved 
at small scales if the density-weighted fluid velocity, $\rho^{1/3} \uu$, is used. 
\end{abstract}
\maketitle

\paragraph*{Introduction.}
Fully developed turbulence is often seen as the last great unsolved problem in classical physics which has 
evaded physical understanding for many decades. Although significant advances have been made in the 
regime of wave turbulence for which a systematic analysis is possible \cite{WT}, the regime of strong 
turbulence -- the subject of this Letter -- continues to resist modern efforts at solution; for that reason any 
exact result is of great importance. In his third 1941 turbulence paper Kolmogorov derived an exact relation 
for incompressible isotropic hydrodynamics in terms of third-order longitudinal structure function and in 
the asymptotic limit of a high Reynolds number ($Re$) \cite{K41}. 
Because of the rarity of such results, the Kolmogorov's universal four-fifths law has a cornerstone role in the 
analysis of turbulence \citep{frisch}. Few extensions of such results to other fluids have been made; 
it concerns for example scalar passively advected such as the temperature or a pollutant in the atmosphere, 
quasi-geostrophic flows or astrophysical magnetized fluids described in the framework of (Hall) MHD \cite{PP98}.  
It is only recently that an attempt to generalize such laws to axisymmetric turbulence has been made but an 
additional assumption is made about the foliation of the correlation space \cite{galtier09}.

The previous results are found for incompressible fluids and to our knowledge no universal law has been 
derived for compressible turbulence (except for the wave turbulence regime \cite{acous}) which is far more 
difficult to analyze. The lack of knowledge is such that even basic statements about turbulence like the presence 
of a cascade, an inertial range and constant flux energy spectra are not well documented \cite{chandra}. 
That is in contrast with the domain of application of compressible turbulence which ranges from aeronautical 
engineering to astrophysics \cite{SC,sagaut,scalo}. In the latter case, it is believed that highly compressible 
turbulence controls star formation in interstellar clouds \cite{astro} whereas in the former case $Re$ is relatively smaller. 

In that context, the pressure-less hydrodynamics is an interesting model to investigate the limit of high Mach number 
compressible turbulence whose simplest form is the one dimension Burgers equation which has been the subject of 
many investigations \cite{bec}. Among the large number of results, we may note that with exact field-theoretical 
methods it is possible to find explicit forms of some probability distributions \cite{polyakov}; it is also possible to derive 
the corresponding exact Kolmogorov law for the third-order structure function \cite{frisch}. 

In the general case, our knowledge of compressible hydrodynamic turbulence is mainly limited to direct numerical 
simulations \citep{pouquet}. The most recent results for supersonic isothermal turbulence with a grid resolution up to 
$2048^3$ \cite{kritsuk} reveal that the inertial range velocity scaling deviates substantially from the incompressible 
Kolmogorov spectrum with a slope of the velocity power spectrum close to $-2$ and an exponent of the third-order 
velocity structure function of about $1.3$. Surprisingly, the incompressible predictions are shown to be restored 
if the density-weighted fluid velocity, $\rho^{1/3} \uu$, is used instead of simply the velocity $\uu$. Although a 
$-2$ spectrum may be associated with shocks -- like in one dimension -- it seems that their contribution in three 
dimensions (3D) is more subtle. Generally speaking it is fundamental to establish the equivalent of the $4/5$s 
law for compressible turbulence before going to the more difficult problem of intermittency \citep{pouquet2}. 

In this Letter, compressible isothermal hydrodynamic turbulence is analyzed in the limit of high $Re$.
We shall investigate the nature of such a compressible turbulence through an analysis in the physical 
space in terms of two-point correlation functions. In particular the discussion is focused on the isotropic case for 
which a simple exact relation emerges. The theoretical predictions illuminate some recent high-resolution direct 
numerical simulations made in the astrophysical context.

\paragraph*{Homogeneous compressible turbulence.}
We start our analysis with the following 3D compressible equations \cite{landau}
\begin{eqnarray}
\p_t \rho + \pmbmath{\nabla} \cdot (\rho \uu ) &=& 0 \, , \nonumber \\
\partial_t (\rho \uu )+ \pmbmath{\nabla} \cdot (\rho \uu \uu) &=& - \pmbmath{\nabla} P 
+ \mu \Delta \uu + {\mu \over 3} \pmbmath{\nabla} (\pmbmath{\nabla} \cdot \uu) + {\bf f} \, , \nonumber 
\label{hd1}
\end{eqnarray}
where $\rho$ is the density, $\uu$ the velocity, $P$ the pressure, $\mu$ the coefficient of viscosity and 
${\bf f}$ a stationary homogeneous external force acting at large scales. The system is closed with the 
isothermal equation $P= C^2_s \rho$ where $C_s$ is the speed of sound. The energy equation takes the form
\be
\p_t \langle E \rangle = - \mu \langle (\nabla \times \uu)^2 \rangle 
- {4 \over 3} \mu \langle (\nabla \cdot \uu)^2 \rangle + F \, , 
\label{energy}
\ee
with $\langle \, \rangle$ an ensemble average (which is equivalent to a spatial average in homogeneous 
turbulence), $E=\rho u^2 / 2 + \rho e$ the total energy, $e=C_s^2 \ln(\rho/\rho_0)$ ($\rho_0$ is a constant 
density introduced for dimensional reasons) and $F$ the energy injected. 

The relevant two-point correlation functions associated with the total energy may be obtained by noting that for 
homogeneous turbulence 
\begin{eqnarray}
\langle \delta (\rho \uu) \cdot \delta \uu \rangle &=&  
2 \langle \rho u^2 \rangle - \langle (\rho + \rho') \uu \cdot \uu' \rangle \, , \\
\langle \delta \rho \, \delta e \rangle &=&  2 \langle \rho e \rangle - \langle \rho e' + \rho' e \rangle \, , 
\end{eqnarray}
where for any variable $\xi$, $\delta \xi \equiv \xi({\bf x} + \rr) - \xi ({\bf x}) \equiv \xi'-\xi$. Then, we find
\be
{{\cal R}(\rr) + {\cal R}(-\rr) \over 2} = \langle E \rangle - {1 \over 4} \langle \delta (\rho \uu) \cdot \delta \uu \rangle 
- {1 \over 2} \langle \delta \rho \delta e \rangle \, , 
\label{corr2}
\ee
where ${\cal R}(\rr) \equiv \langle \rho \uu \cdot \uu' / 2 + \rho e' \rangle \equiv \langle R \rangle$ 
and ${\cal R}(-\rr) \equiv \langle \rho' \uu' \cdot \uu / 2 + \rho' e \rangle \equiv \langle {\tilde R} \rangle$. Note that 
for homogeneous compressible turbulence the relation, ${\cal R}(\rr) = {\cal R}(-\rr)$, holds only when isotropy 
is assumed whereas it is always valid in the incompressible limit for which ${\cal R}$ is reduced to a second-order 
velocity correlation function \citep{batch67}. As we will see below, relation (\ref{corr2}) is very helpful for deriving 
an exact relation for some two-point correlation functions. In practice, we shall derive a dynamical equation for 
${\cal R}(\rr)$; first, we have to compute 
\begin{eqnarray}
\p_t \langle \rho \uu \cdot \uu' \rangle &=& \langle \rho \uu \cdot \p_t \uu' + \uu' \cdot \p_t (\rho \uu) \rangle 
\nonumber \\
&=& \langle \rho \uu \cdot ( - \uu' \cdot \pmbmath{\nabla}' \uu' - {1 \over \rho'} \pmbmath{\nabla}' P') \rangle 
\label{toti} \\
&& + \langle \uu' \cdot (- \pmbmath{\nabla} \cdot ( \rho \uu \uu ) - \pmbmath{\nabla} P ) \rangle 
+ 2{\cal D} + 2{\cal F} \, , \nonumber
\label{n1a} 
\end{eqnarray}
where for simplicity $2{\cal D}$ and $2{\cal F}$ denote respectively the contributions to the correlation of the 
viscous and forcing terms. By remarking that
\be
\langle {\rho \over \rho'} \uu \cdot \pmbmath{\nabla}' P' \rangle = 
\langle \rho C_s^2 u_\ell {\p_\ell' \rho' \over \rho'} \rangle \nonumber 
= \langle \rho u_\ell \p_\ell' e' \rangle 
= \langle \pmbmath{\nabla}' \cdot (\rho e' \uu) \rangle \, , \nonumber
\ee
and that
\be
\langle \uu' \cdot \pmbmath{\nabla}'(\rho \uu \cdot \uu') \rangle = 
\langle \pmbmath{\nabla}' \cdot (\rho (\uu \cdot \uu') \uu') - 
\rho (\uu \cdot \uu') (\pmbmath{\nabla}' \cdot \uu') \rangle \, , \nonumber
\ee
we can rewrite (\ref{toti}) in the following way
\begin{eqnarray}
\p_t \langle \rho \uu \cdot \uu' \rangle &=& \langle - \uu' \cdot \pmbmath{\nabla}' (\rho \uu \cdot \uu') 
- \pmbmath{\nabla}' \cdot (\rho e' \uu) \rangle \nonumber \\
&& - \langle \pmbmath{\nabla} \cdot ( \rho (\uu \cdot \uu') \uu  + P \uu' ) \rangle + 2{\cal D} + 2{\cal F}\nonumber \\
&=& \pmbmath{\nabla}_\rr \cdot \langle - \rho (\uu \cdot \uu') \delta \uu + P \uu' - \rho e' \uu \rangle \nonumber \\
&& + \langle \rho (\uu \cdot \uu') (\pmbmath{\nabla}' \cdot \uu') \rangle + 2{\cal D} + 2{\cal F} \, . 
\label{n1}
\end{eqnarray}
Secondly, we have to complete the computation with 
\begin{eqnarray}
\p_t \langle \rho e' \rangle &=& \langle \rho \p_t e' + e' \p_t \rho \rangle = 
\langle C_s^2 {\rho \over \rho'} \p_t \rho' + e' \p_t \rho \rangle \nonumber \\
&=& \langle - C_s^2 {\rho \over \rho'} \pmbmath{\nabla}' \cdot (\rho' \uu') - 
e' \pmbmath{\nabla} \cdot (\rho \uu) \rangle \nonumber \\
&=& - \langle \pmbmath{\nabla}' \cdot (C_s^2 \rho \uu') \rangle + 
\langle \rho' \uu' \cdot \pmbmath{\nabla}' (C_s^2 {\rho \over \rho'}) \rangle 
\nonumber \\
&& - \langle \pmbmath{\nabla} \cdot (\rho e' \uu) \rangle \, .
\label{n2a}
\end{eqnarray}
By noting that
\be
\langle \rho' u'_\ell \p'_\ell  (C_s^2 {\rho \over \rho'}) \rangle 
= - \langle \pmbmath{\nabla}' \cdot ( \rho e' \uu') \rangle + 
\langle e' \pmbmath{\nabla}' \cdot (\rho \uu') \rangle \, , \nonumber 
\ee
we obtain after simplification 
\begin{eqnarray}
\p_t \langle \rho e' \rangle &=& 
\pmbmath{\nabla}_\rr \cdot \langle - \rho e' \delta \uu - P \uu' \rangle + 
\langle \rho e' (\pmbmath{\nabla}' \cdot \uu') \rangle \, . 
\label{n2}
\end{eqnarray}
The combination of (\ref{n1}) and (\ref{n2}) leads to 
\begin{eqnarray}
\p_t {\cal R}(\rr) &=& \langle (\pmbmath{\nabla}' \cdot \uu') R \rangle + {\cal D} + {\cal F} \label{n3} \\
&&+ \pmbmath{\nabla}_\rr \cdot \left\langle - R \, \delta \uu
- {1 \over 2} P \uu' - {1 \over 2} \rho e' \uu \right\rangle \, . \nonumber
\end{eqnarray}
The same type of analysis may be performed for ${\cal R}(-\rr)$ which eventually leads to the dynamical equation
\begin{eqnarray}
&&\p_t \left({{\cal R}(\rr) + {\cal R}(-\rr) \over 2} \right) = \label{n4} \\
&&{1 \over 2} \langle (\pmbmath{\nabla}' \cdot \uu') R \rangle + 
{1 \over 2} \langle (\pmbmath{\nabla} \cdot \uu) {\tilde R} \rangle
+ {1 \over 2} ({\cal D} + {\cal {\tilde D}} + {\cal F} + {\cal {\tilde F}}) \nonumber \\
&& + {1 \over 2} \pmbmath{\nabla}_\rr \cdot \left\langle - ( R+{\tilde R} ) \delta \uu
- {1 \over 2} (P \uu' - \rho e' \uu - P' \uu + \rho' e \uu') \right\rangle \, , \nonumber
\end{eqnarray}
where ${\cal {\tilde D}}$ and ${\cal {\tilde F}}$ denote respectively the additional contribution of the viscous 
and forcing terms.

\paragraph*{Local turbulence.}
For the final step of the derivation we shall introduce the usual assumption specific to 3D fully 
developed turbulence with a direct energy cascade \citep{frisch,landau}. In particular, we suppose the 
existence of a statistical steady state in the infinite Reynolds number limit with a balance between forcing and 
dissipation. We recall that the dissipation is a sink for the total energy and acts mainly at the smallest scales of 
the system. Then, far in the inertial range we may neglect the contributions of ${\cal D}$ and ${\cal {\tilde D}}$ 
in equation (\ref{n4}) \citep{aluie}. The introduction of structure functions leads to the final form
\begin{eqnarray}
&-& 2\varepsilon = \langle (\pmbmath{\nabla}' \cdot \uu') (R-E) \rangle 
+ \langle (\pmbmath{\nabla} \cdot \uu) ({\tilde R}-E') \rangle  \label{iso1} \\
&+& \pmbmath{\nabla}_\rr \cdot \left\langle 
\left[ {\delta (\rho \uu) \cdot \delta \uu \over 2} + \delta \rho \delta e - C_s^2 {\bar \delta} \rho \right] \delta \uu 
+ {\bar \delta} e \delta (\rho \uu) \right\rangle , \nonumber 
\end{eqnarray}
where ${\bar \delta} X \equiv (X+X')/2$ and $\varepsilon$ is the mean total energy injection rate (which is equal 
to the mean total energy dissipation rate; see relation (\ref{energy})). 
Note that at relatively small $r$ the function, $R-E$ (and ${\tilde R}-E'$), and its derivative are negative since 
the correlation between two points is maximum if the points are the same. 

It is straightforward to show that in the limit of incompressible turbulence we recover the well-known 
expression; indeed we obtain (with $\rho \to \rho_0 = 1$)
\begin{eqnarray}
- 4 \varepsilon = \pmbmath{\nabla}_\rr \cdot  \langle (\delta \uu)^2 \delta \uu \rangle \, ,
\label{incomp}
\end{eqnarray}
which is the primitive form of the Kolmogorov's law. An integration over a ball of radius $r$ leads to the 
well-known expression \citep{antonia}, $- (4/3) \varepsilon r = \langle (\delta \uu)^2 \delta u_r \rangle$, 
where $r$ means the radial (often called longitudinal) component, \ie the one along the direction $\rr$.

\paragraph*{Isotropic turbulence.}
Expression (\ref{iso1}) is the main result of the Letter. It is an exact relation for some two-point correlation functions 
when fully developed turbulence is assumed. It is valid for homogeneous -- non necessarily isotropic -- 3D 
compressible isothermal turbulence. 
Note that the pressure contribution appears through the term $C_s^2 {\bar \delta}\rho$ and is therefore 
negligible in the large Mach number limit ($C_s \to 0$).
When isotropy is additionally assumed this relation can be written symbolically as 
\be
- 2 \varepsilon = {\cal S}(r) + {1 \over r^2} \p_r(r^2 {\cal F}_r) \, , 
\label{iso1z}
\ee
where ${\cal F}_r$ is the radial component of the isotropic energy flux vector. In comparison with the 
incompressible case (\ref{incomp}), expression (\ref{iso1z}) reveals the presence of a new type of term 
${\cal S}$ which is by nature compressible since it is proportional to the dilatation (\ie the divergence of 
the velocity). This term has a major impact on the nature of compressible turbulence since as we will see 
it acts like a source or a sink for the mean energy transfer rate. Note that ${\cal S}$ consists of two terms 
which account for two-point measurement approach.

\paragraph*{Discussion.}
We may further reduce equation (\ref{iso1z}) by performing an integration over a ball of radius $r$. After 
simplification we find the exact relation 
\be
- {2 \over 3} \varepsilon r = {1 \over r^2} \int_0^r {\cal S}(r) r^2 dr + {\cal F}_r(r) 
\label{THI} \, .
\ee

We start the discussion by looking at the small scale limit of the previous relation which means that the 
scales are assumed to be small enough to perform a Taylor expansion but not too small to be still in the 
inertial range. We obtain ${\cal S}(r) = {\cal S}(0) + r \p_r {\cal S}(0)=r \p_r {\cal S}(0)$ which leads to
\be
- {2 \over 3} \left[ \varepsilon + {3 \over 8} r \p_r {\cal S}(0) \right] r \equiv - {2 \over 3} \varepsilon_{\rm{eff}} r 
=  {\cal F}_r(r) \, .
\label{THI2}
\ee
Note that we do not assume the cancellation of the first derivative of $S$ at $r=0$ although the function, 
$R-E$, reaches an extremum at $r=0$; the reason is that this function is weighted by the dilatation function 
which may have a non trivial form. 
We see that at the leading order the main contribution of ${\cal S}(r)$ is to modify $\varepsilon$ for giving 
an effective mean total energy injection rate $\varepsilon_{\rm{eff}}$. Then, the physical interpretation of 
(\ref{THI2}) is the following. When the flow is mainly in a phase of dilatation (positive velocity divergence), 
the additional term is negative and $\varepsilon_{\rm{eff}}$ is smaller than $\varepsilon$. On the contrary, in 
a phase of compression $\p_r {\cal S}(0)$ is positive and $\varepsilon_{\rm{eff}}$ is larger than $\varepsilon$. 
\begin{figure}[ht]
\resizebox{150mm}{!}{\includegraphics{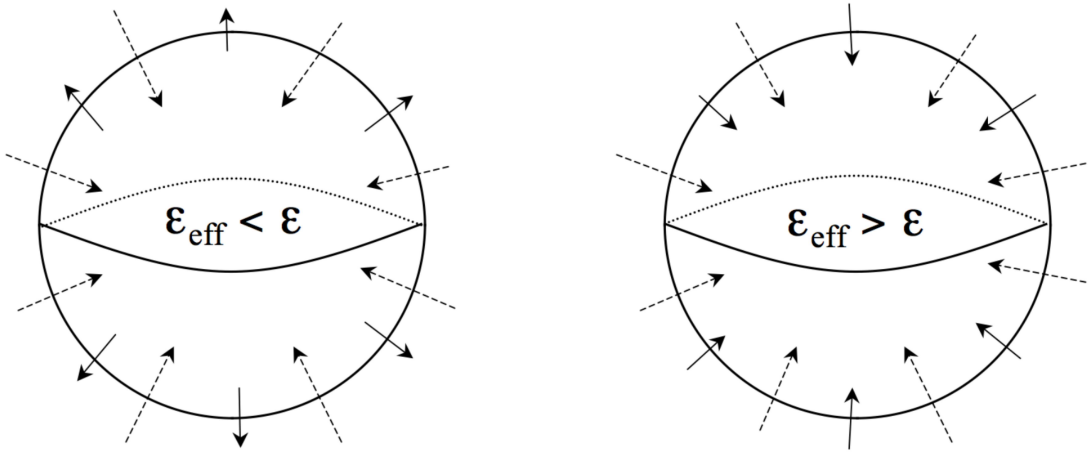}}
\caption{Dilatation (left) and compression (right) phases in space correlation for isotropic turbulence. In a direct 
cascade scenario the flux vectors $\pmbmath{\cal F}$ (dashed arrows) are oriented towards the center of the 
sphere. Dilatation and compression (solid arrows) are additional effects which act respectively in the opposite 
or in the same direction as the flux vectors. 
\label{Fig1}}
\end{figure}

An illustration of dilation and compression effects in the space correlation is given in Fig. \ref{Fig1}. In both cases, 
the flux vector $\pmbmath{\cal F}$ (dashed arrows) is oriented towards the center of the sphere ($r=0$) since a 
direct cascade is expected. Dilatation and compression act additionally (solid arrows): in the first case, the effect 
is similar to a decrease of the local mean total energy transfer rate whereas in the second case it is similar to an 
increase of the local mean total energy transfer rate. 

The discussion may be extended to the entire inertial range (\ie for larger values of $r$)
when the (turbulent) Mach number is relatively high. In this case the analysis 
is focused on expression (\ref{THI}) for which we have already noted that a term like $R-E$ is mainly negative. 
It is interesting to note that ${\cal S}(r)$ is composed of two types of term which are different by nature. First, 
there is the dilatation dominated by the smallest scales in the flow -- the shocklets -- which mainly give a negative 
contribution with a fast variation \citep{smith}. Secondly, there is the correlation $R-E$ which derives most of its 
contribution from relatively larger scales with a slower variation. This remark may lead to the assumption that both 
terms are relatively decorrelated \cite{aluie}. Then ${\cal S}(r)$ may be simplified as (by using relation (\ref{corr2}))  
\be
{\cal S}(r) \simeq - \left \langle {\bar \delta} (\pmbmath{\nabla} \cdot \uu) 
\left[{1 \over 4} \delta (\rho \uu) \cdot \delta \uu + {1 \over 2} \delta \rho \delta e \right] \right\rangle \, .
\label{approx}
\ee
The previous expression is not derived rigorously but it may give us some intuition about its contribution. 
For example, we may expect a power law dependence close to $r^{2/3}$ for the structure functions. Direct 
numerical simulations have never shown a scale dependence for the dilatation and we may expect that 
it behaves like a relatively small factor. Then ${\cal S}(r)$ will still modify $\varepsilon$ as explained in the 
discussion above, however the power law dependence in $r$ would be now slightly different. 
In conclusion and according to this simple analysis we see that compression effects (through the dilatation) 
will mainly impact the scaling law at the largest scales.

\paragraph*{Compressible spectrum.}
We may try to predict a power law spectrum for compressible turbulence. First, we note that several 
predictions have been made for the kinetic energy spectrum and also for the spectra associated with the 
solenoidal or the compressible part of the velocity \citep{kadomtsev}. We recall that although these 
decompositions are convenient for analytical developments, the associated energies are not inviscid 
invariants and the predictions are heuristic. For incompressible turbulence the situation is different because 
a prediction in $k^{-5/3}$ for the kinetic energy spectrum may be proposed by applying a dimensional 
analysis directly on the $4/5$s law \cite{frisch}. Although it is not an exact prediction, the $4/5$s law gives a 
stronger foundation to the energy spectrum for which a constant flux is expected. This remark was already 
noted in particular in recent 3D direct numerical simulations of isothermal turbulence where it is observed that 
the Kolmogorov scaling is not preserved for the spectra based only on the velocity fluctuations \citep{kritsuk}. 

We shall derive a power law spectrum for compressible turbulence by applying a dimensional analysis 
on equation (\ref{iso1}). Dimensionally, we may find $\varepsilon_{\rm{eff}} r \sim \rho u^3$. By introducing 
the density-weighted fluid velocity, ${\bf v} \equiv \rho^{1/3} \uu$, and following Kolmogorov we obtain 
$E^v(k) \sim \varepsilon_{\rm{eff}}^{2/3} k^{-5/3}$,
where $E^v(k)$ is the spectrum associated to the variable ${\bf v}$. Our prediction is compatible with the 
measurements recently made by direct numerical simulations \citep{kritsuk} where the authors have noted 
that the exponent of the third-order velocity structure function is close to one if the field used is ${\bf v}$ 
instead of $\uu$. (Note that two other scaling relations may be predicted like for the pressure term.) 
As explained by several authors \citep{kadomtsev} in compressible turbulence we do not expect a constant 
flux in the inertial range. Here, the same conclusion is reached since we are dealing with an effective mean 
energy transfer rate. More precisely if we expect a power law dependence in $k$ for the effective transfer 
rate one arrives at the conclusion that a steeper power law spectrum may happen at the largest scales. 
According to relation (\ref{approx}) and the simple estimate, $\rho v^2 \sim r^{2/3}$, we could have 
$E^v(k) \sim k^{-19/9}$. This prediction means that for a small prefactor in (\ref{approx}) one needs an extended 
inertial range to feel the compressible effects on the power spectrum. 
The scale at which the transition happens between $-19/9$ and $-5/3$ may be the sonic scale $k_s$ as 
proposed in \cite{federrath} where such power laws were detected; in our case, a rough estimate gives 
$k_s \sim \langle (\nabla \cdot \uu) / \delta \uu \rangle$.

\paragraph*{Conclusion.}
The present work opens important perspectives to further understand the nature of compressible turbulence 
in the asymptotic limit of large Reynolds numbers with the possibility to extend the analysis to magnetized 
fluids with possibly other types of closures (\eg polytropic gas), or to improve intermittency models by using the new 
relation -- obtained by a statistical analysis at low order -- as pivotal for a heuristic extension to statistical laws at 
higher order. We believe that astrophysics (\eg interstellar turbulence) is one of the most important domain 
of application of the present work \citep{scalo}.

\paragraph*{Acknowledgment.}
We acknowledge S. Boldyrev, A. Kritsuk and T. Passot for useful discussions.

\end{document}